\documentclass{PoS}
\usepackage{cite}
\usepackage{amsmath}
\newcommand{\ber}{\begin{eqnarray}} 
\newcommand{\eer}{\end{eqnarray}}

\title{How well do we know neutrino-electron scattering? EFT approach}

\ShortTitle{How well do we know neutrino-electron scattering? EFT approach}

\author{Oleksandr Tomalak \thanks{Preprint number: FERMILAB-CONF-19-562-T}\\
       Department of Physics and Astronomy, University of Kentucky, Lexington, KY 40506, USA\\
       Fermilab, Batavia, IL 60510, USA\\
       E-mail: \email{oleksandr.tomalak@uky.edu}}

\abstract{We determine the effective theory of neutrino-electron and neutrino-quark scattering and provide the most precise up-to-date prediction for neutrino-electron scattering cross sections quantifying errors for the first time to be of order $0.2-0.4~\%$. Radiative corrections in the theory with electron and neutrinos only involve three effective couplings. One is the Fermi constant which is known with sub-ppm accuracy. Another one has a small error of order $0.02~\%$. The uncertainty of the third one is limited by the knowledge of hadronic contributions to charge-isospin vector-vector correlation function.}

\FullConference{The 21st international workshop on neutrinos from accelerators (NuFact2019)\\
		August 26 - August 31, 2019\\
		Daegu, Korea}

\begin{document}

Neutrino-electron scattering is an attractive process for the neutrino physics community. 
Historically, it gave us confirmation of weak neutral currents and first measurements in developing the Standard Model 
of particle physics. The process plays an important role in studies of solar neutrinos
and reactor antineutrinos. Moreover, it provides a tool to constrain the neutrino flux at accelerator-based experiments
that is free from nuclear uncertainties~\cite{Park:2015eqa,Valencia:2019mkf,Marshall:2019vdy}.

As an exactly calculable reaction, neutrino-electron scattering has attracted a lot of attention.
The leading-order Lagrangian and unpolarized cross sections were obtained in pioneering works of Weinberg and `t Hooft~\cite{Weinberg:1967tq,tHooft:1971ucy}.
Afterwards, electroweak~\cite{Marciano:1980pb,Aoki:1980ix,Sarantakos:1982bp} and QED corrections with one-photon bremsstrahlung~\cite{Sarantakos:1982bp,Lee:1964jq,Ram:1967zza,Hioki:1981gi,Bardin:1985fg,Passera:2000ug} 
were evaluated by numerous authors. 
However, the first consistent effective field theory calculation of this process providing an error estimate has appeared just recently~\cite{Tomalak:2019ibg}.
In this work, we aim to present a complete description of neutrino-electron scattering at sub-percent
level of accuracy and complement the picture with neutrino-quark interaction.

\begin{figure}[htb]
\begin{center}
\vspace{-0.25cm}	
\hspace{0.7cm}	\includegraphics[scale=1.]{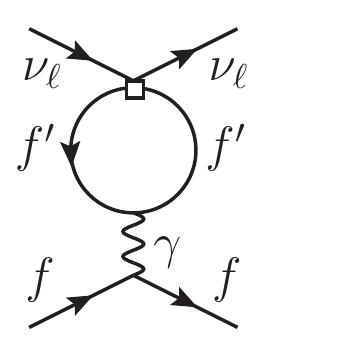}\hspace{0.cm}
\end{center}
	\caption{The leading contribution to running of couplings in effective theory.\label{one_loop_ET}}
\end{figure}

The typical momentum transfer in the elastic neutrino-electron scattering process $Q^2$ is bounded from above, $Q^2 < 2 m_e\mathrm{E_\nu}$, where $\mathrm{E_\nu}$ is the incoming neutrino energy and $m_e$ is the electron mass. The neutrino flux in accelerator-based experiments is peaked at relatively low energies around $\mathrm{E_\nu} \simeq 0.5-3~\mathrm{GeV}$. Neutrino scattering with $\mathrm{E_\nu} \lesssim 10~\mathrm{GeV}$ corresponds to $Q^2 \lesssim 0.01~\mathrm{GeV}^2$ and is not directly sensitive to hadron and quark dynamics. Therefore, all physics in the low-energy neutrino-electron scattering as well as decay of muon can be accurately described by electron, muon and neutrino degrees of freedom in an effective QED field theory with contact four-fermion operators~\cite{Fermi:1934hr,Feynman:1958ty,Sudarshan:1958vf}. The effective four Fermi Lagrangian ${\cal L}_\mathrm{F}$ is given by
\ber
{\cal L}_\mathrm{F} &=& - \sum_{\substack{\ell= e, \mu, \tau \\ \ell' = e,\mu}}   \bar{\nu}_\ell \gamma^\sigma \mathrm{P}_\mathrm{L} \nu_\ell
    \, \bar{\ell}' \gamma_\sigma (c_\mathrm{L}^{\nu_\ell \ell'} \mathrm{P}_\mathrm{L}
    + c_\mathrm{R}^{\nu_\ell \ell'} \mathrm{P}_\mathrm{R}) \ell' 
    - c \left( 
  \bar{\nu}_{\mu }\gamma^\sigma \mathrm{P}_\mathrm{L} \nu_{e}
  \, \bar{e} \gamma_\sigma \mathrm{P}_\mathrm{L}  \mu + \mathrm{h.c.} \right) \,. \label{effective_Lagrangian}
\eer
Here $\mathrm{P}_\mathrm{L}$ and $\mathrm{P}_\mathrm{R}$ are projectors on the left-handed and right-handed chiral states respectively:
\ber
\mathrm{P}_\mathrm{L} = \frac{1-\gamma_5}{2}, \qquad \mathrm{P}_\mathrm{R} = \frac{1+\gamma_5}{2}.
\eer
 $e,~\mu$ and $\nu_\ell$ denote electron, muon and corresponding neutrino fields, and $c^{\nu_\ell \ell'}_\mathrm{L,R},~c$ are effective couplings. We determine effective couplings by matching the effective theory to the Standard Model at the electroweak scale~\cite{Tanabashi:2018oca} through order $\mathrm{O} \left(\alpha \alpha_s\right)$ in $\overline{\rm MS}$ renormalization scheme with a subsequent running to low energies. The running is governed by the closed loop contributions in Fig.~\ref{one_loop_ET}, where all degrees of freedom in the theory appear in the loop.
 
Couplings $c^{\nu_\ell e}_\mathrm{R}$ depend on the scale $\mu$ within $\overline{\mathrm{MS}}$ renormalization scheme.
Others can be determined from scale-independent combinations
\ber
c_\mathrm{L}^{\nu_\tau e} \left( \mu \right)- c^{\nu_\tau e}_\mathrm{R} \left( \mu \right)  = c_\mathrm{L}^{\nu_\mu e} \left( \mu \right) - c^{\nu_\mu e}_\mathrm{R} \left( \mu \right) &=&  - \sqrt{2} \mathrm{\tilde{G}}_e, \\
c \left( \mu \right) =  2 \sqrt{2} \mathrm{G}_\mathrm{F}, \qquad c_\mathrm{L}^{\nu_e e} \left( \mu \right)  - c^{\nu_e e}_\mathrm{R} \left( \mu \right)&=&   - \sqrt{2} \mathrm{\tilde{G}}_e + 2 \sqrt{2} \mathrm{G}_\mathrm{F}, 
\eer
with the Fermi coupling $ \mathrm{G}_\mathrm{F}$ and a constant $\mathrm{\tilde{G}}_e$:
\ber
 \mathrm{G}_\mathrm{F} &=& 1.1663787(6) \times 10^{-5}~\mathrm{GeV}^{-2}, \qquad
 \mathrm{\tilde{G}}_e = 1.18083(21) \times 10^{-5}~\mathrm{GeV}^{-2}.
\eer
The uncertainty of the latter is mainly from neglected higher order perturbative corrections (estimated by varying the matching scale by a factor $\sqrt{2}$) with
a subdominant error from input parameters.

Above the $\tau$-mass scale, the right-handed coupling is flavor independent: 
\ber
c^{\nu_\tau e}_\mathrm{R} \left( \mu \right) =c^{\nu_\mu e}_\mathrm{R} \left( \mu \right) = c^{\nu_e e}_\mathrm{R} \left( \mu \right) = c_\mathrm{R} \left( \mu \right), \qquad \mu \geq m_\tau.
\eer
For neutrino-electron scattering and muon decay applications, radiative corrections can be evaluated in the leptonic theory. At scale 
$\mu = 2~\mathrm{GeV}$ in $\overline{\mathrm{MS}}$ renormalization scheme, the undetermined constant is~\cite{Tomalak:2019ibg} 
$c_\mathrm{R} \left( \mu =2~\mathrm{GeV} \right) = 0.7773(28) \times 10^{-5}~\mathrm{GeV}^{-2}$, 
with the dominant uncertainty coming from hadronic contributions in Fig.~\ref{one_loop_ET}.
Due to low momentum transfer of neutrino scattering process compared to the hadronic scale, 
this contribution can be evaluated at $Q^2 = 0$ and was integrated out in Refs.~\cite{Tomalak:2019ibg,in_preparation}. Equivalently, radiative corrections can be calculated 
in the theory with electron and neutrinos only with couplings at muon mass scale:
\ber
c^{\nu_e e}_\mathrm{R} \left( m_\mu \right) =c^{\nu_\mu e}_\mathrm{R} \left( m_\mu \right) &=&0.7706(29) \times 10^{-5}~\mathrm{GeV}^{-2}, \\
 c^{\nu_\tau e}_\mathrm{R} \left( m_\mu \right)&=&0.7779(29) \times 10^{-5}~\mathrm{GeV}^{-2},
\eer
and in QED limit:
\ber
c^{\nu_e e}_\mathrm{R} \left( m_e\right)&=&0.7575(29) \times 10^{-5}~\mathrm{GeV}^{-2}, \qquad
c^{\nu_\mu e}_\mathrm{R}\left( m_e\right) = 0.7711(29) \times 10^{-5}~\mathrm{GeV}^{-2}, \\
c^{\nu_\tau e}_\mathrm{R} \left( m_e\right)&=&0.7784(29) \times 10^{-5}~\mathrm{GeV}^{-2},
\eer
which determine oscillations of neutrinos in matter.

\begin{figure}[htb]
          \centering
          \includegraphics[height=0.3\textwidth]{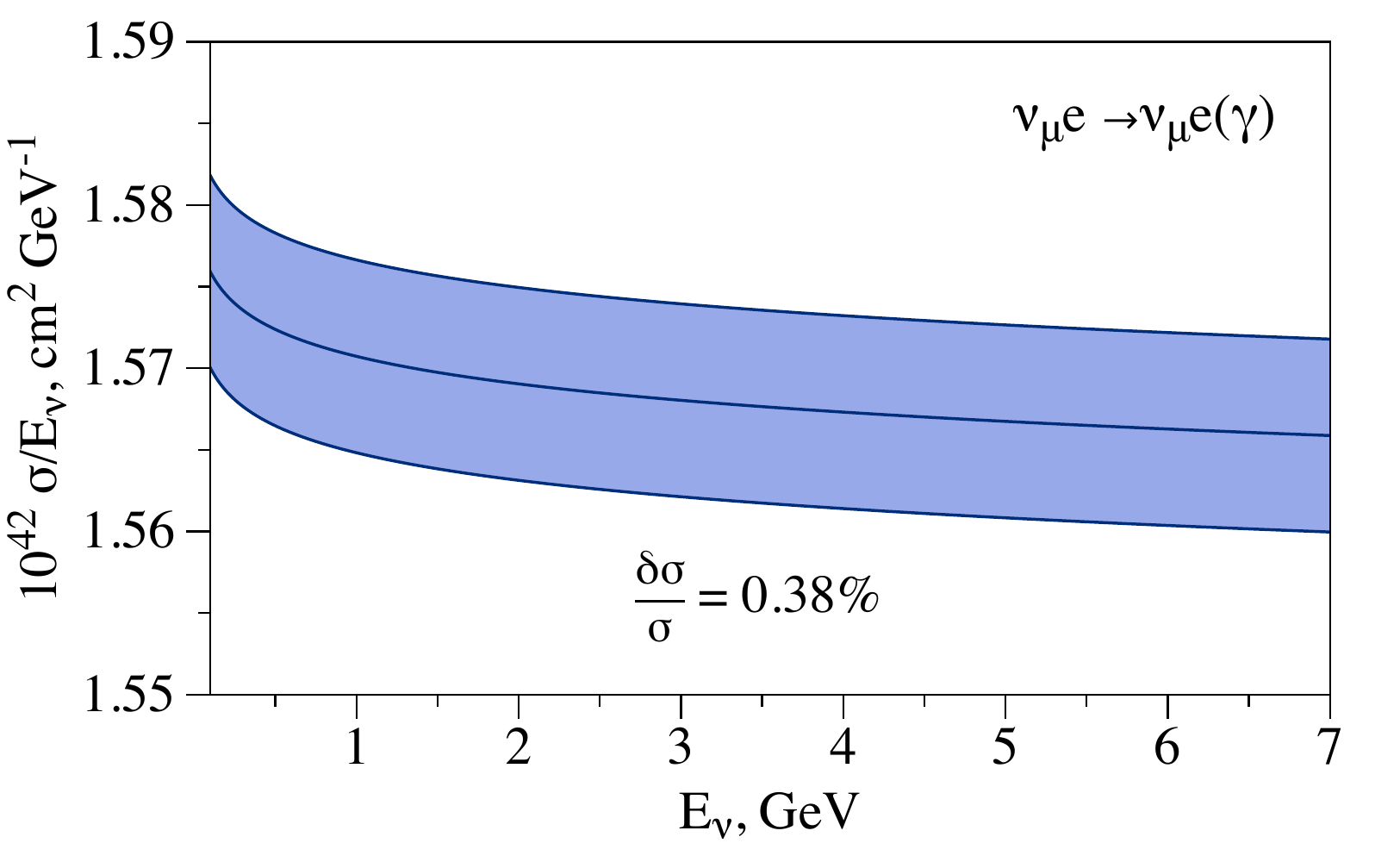}   
          \includegraphics[height=0.3\textwidth]{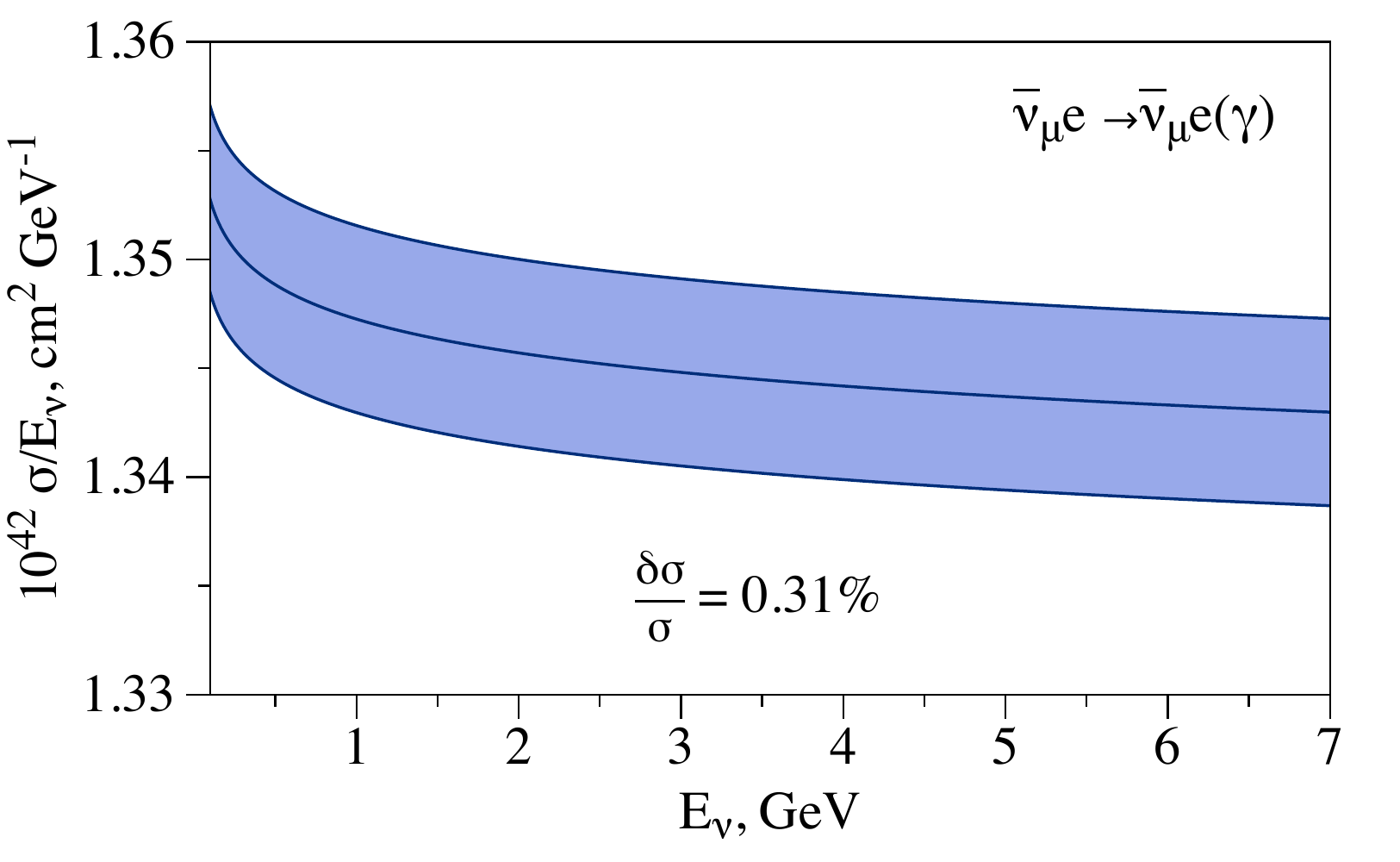}    
          \includegraphics[height=0.3\textwidth]{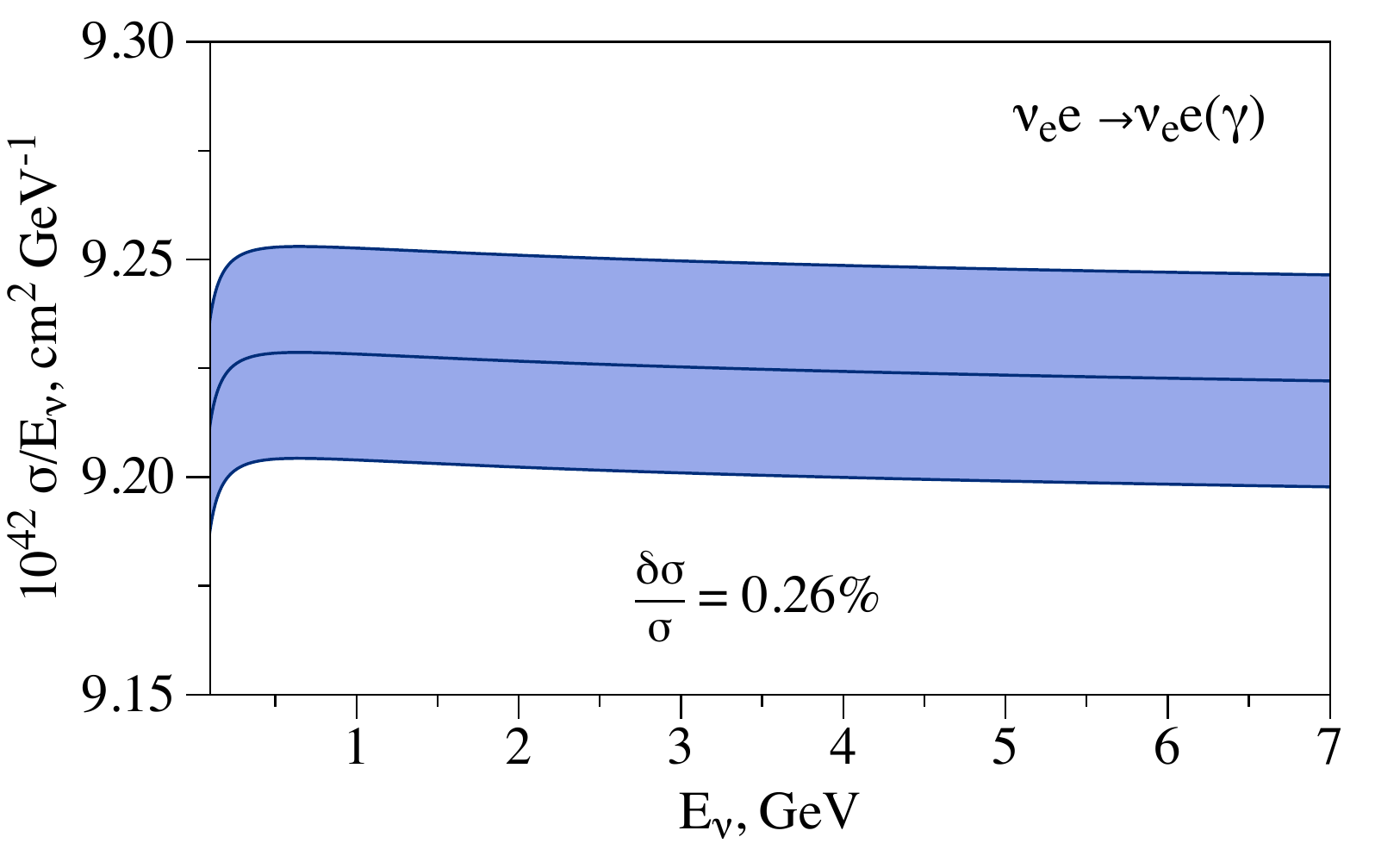}   
          \includegraphics[height=0.3\textwidth]{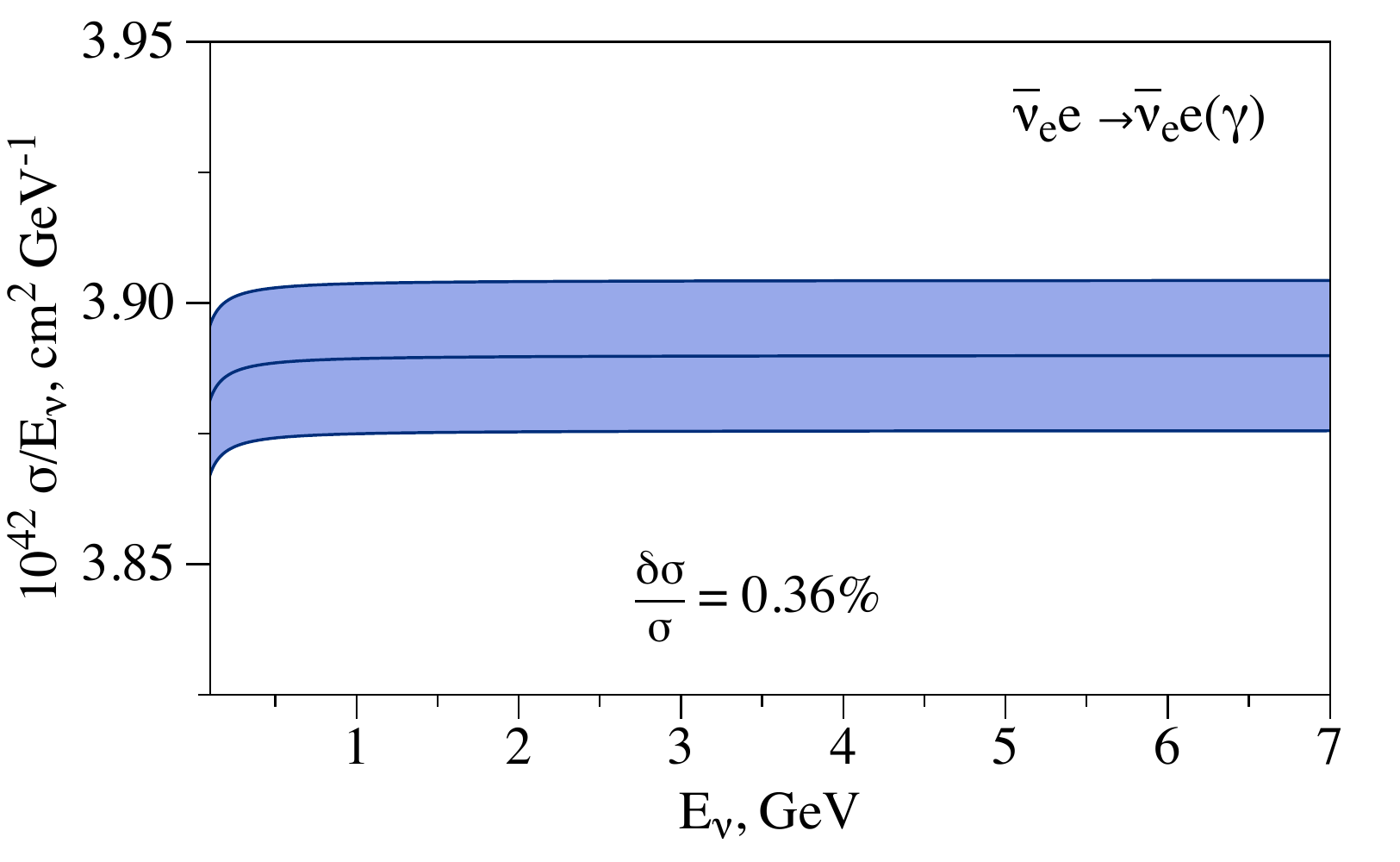}             
          \caption{Total cross section in the (anti-)neutrino-electron
            scattering processes $\nu_\mu e \to \nu_\mu e(
            X_\gamma),~\nu_e e \to \nu_e e(
            X_\gamma),~\bar{\nu}_\mu e \to \bar{\nu}_\mu e(
            X_\gamma)$ and $\bar{\nu}_e e \to \bar{\nu}_e e(
            X_\gamma)$ as a function of (anti-)neutrino beam energy $\mathrm{E_\nu}$. 
            The energy-independent relative uncertainty mainly from 
            the charge-isospin hadronic contribution is also presented.\label{fig:nu_e_delta}}
\end{figure}

Within the effective theory, we evaluate the absolute total cross section for neutrino-electron scattering 
including virtual QED corrections, i.e., vertex and propagator corrections and the closed fermion loop
contribution of Fig.~\ref{one_loop_ET}, and radiation of one real photon~\cite{Tomalak:2019ibg}. 
We present the results for $\nu_\mu e,~\nu_e e,~\bar{\nu}_\mu e$ and $\bar{\nu}_e e$  
scattering in Fig.~\ref{fig:nu_e_delta} and provide an error estimate for the first time. 
The cross section grows linearly with incoming neutrino energy $\mathrm{E_\nu}$
while the relative uncertainty is approximately constant. In Ref.~\cite{Tomalak:2019ibg}, we also evaluate
various spectra, double- and triple-differential distributions both for finite electron mass (applicable to low-energy neutrinos)
and in the limit of small electron mass (for applications to high-energy neutrino beams).

\begin{figure}[htb]
          \centering
          \includegraphics[height=0.48\textwidth]{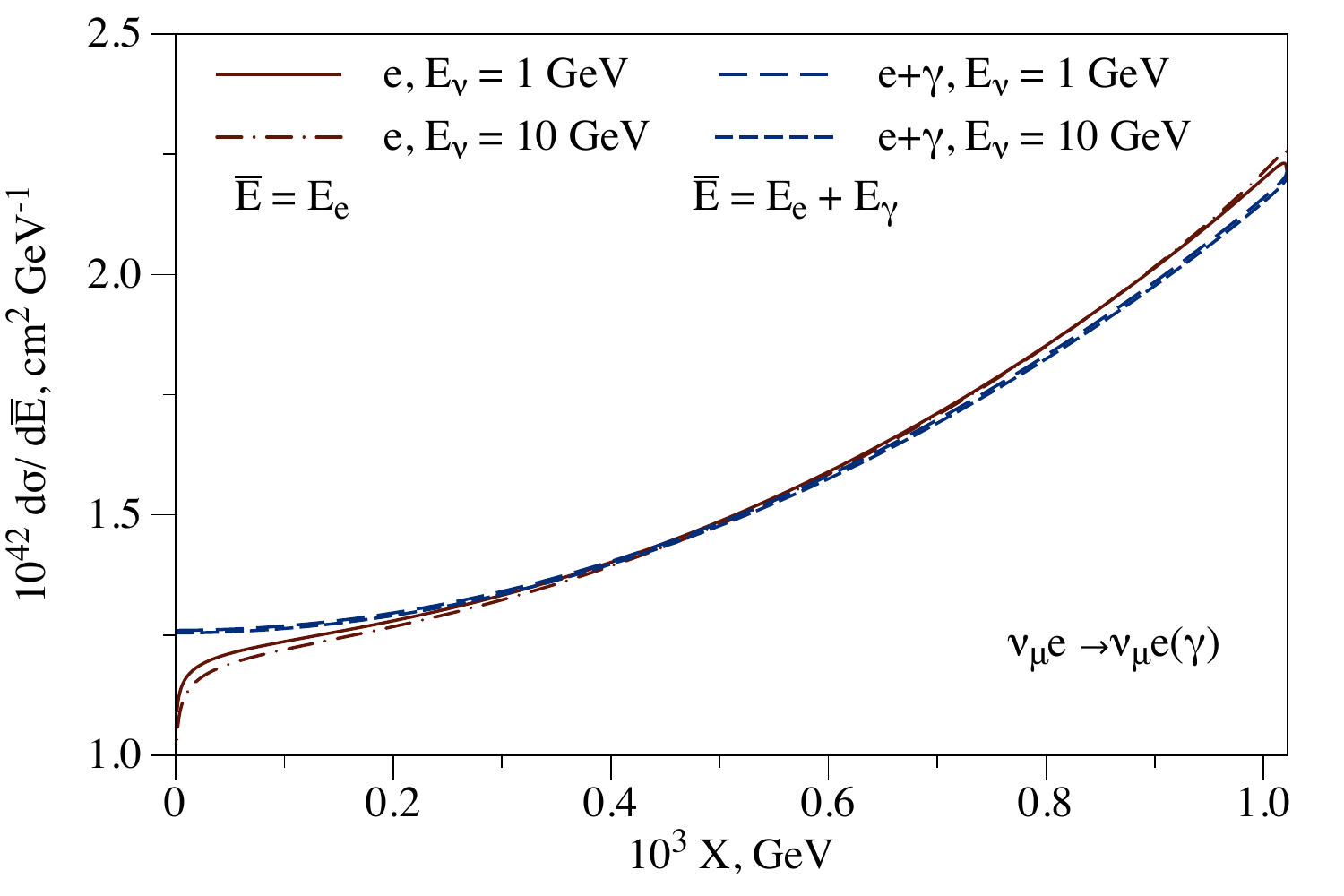}       
          \caption{Energy spectrum in the neutrino-electron
            scattering $\nu_\mu e \to \nu_\mu e(
            \gamma)$, plotted as a function of
            $\mathrm{X}=2m_e(1-\mathrm{\bar{E}}/\mathrm{E_\nu}) $ for two neutrino beam energies $\mathrm{E_\nu} = 1, 10~\,{\rm GeV}$. The solid and dashed-dotted curves correspond
            with electron spectrum, i.e., $\mathrm{\bar{E}} = \mathrm{E_e}$, dashed  curves
            with electromagnetic spectrum, i.e., $\mathrm{\bar{E}} = \mathrm{E_e} + \mathrm{E_\gamma}$.
    \label{fig:nu_e_Etheta2}}
\end{figure}

We stress that bremsstrahlung must be treated carefully, in accordance with experimental conditions.
We concentrate on the scattering 
of muon neutrino flavor for definiteness. In Fig.~\ref{fig:nu_e_Etheta2},
we compare the energy spectrum w.r.t. recoil electron energy $\mathrm{\bar{E}} = \mathrm{E_e}$ to the spectrum w.r.t. sum of electron and photon 
energies $\mathrm{\bar{E}} = \mathrm{E_e} + \mathrm{E_\gamma}$, as a function of the variable $\mathrm{X}$:
\ber
  \mathrm{X} = 2m_e\left( 1- \frac{\mathrm{\bar{E}}}{\mathrm{E_\nu}} \right) \,,
\eer
which becomes $\mathrm{X} \approx \mathrm{E_e} \theta_e^2$ for (anti-)neutrinos of high energy in the case of the electron energy spectrum, where $\theta_e$ is the electron scattering angle. 
Although the integral of both curves is identical, applying an experimental cut on variable $\mathrm{X}$ can lead to an under- or over-estimate of the signal if the chosen distribution does not conform to experimental conditions. This would lead to inaccuracy in neutrino flux calibration.

For completeness, we present also the effective Lagrangian of neutrino-quark interactions ${\cal L}^q_\mathrm{F}$:
\ber
  {\cal L}^q_\mathrm{F}  =    -  \sum_{q,\ell}  \bar{\nu}_\ell \gamma^\mu \mathrm{P}_\mathrm{L} \nu_\ell \,
    \bar{q} \gamma_\mu (c_\mathrm{L}^{q} \mathrm{P}_\mathrm{L} + c_\mathrm{R}^{q} \mathrm{P}_\mathrm{R}) q  - \sum_{q \ne q^\prime, \ell}  \left( c^{q q'} 
  \bar{\ell} \gamma^\mu \mathrm{P}_\mathrm{L} \nu_{\ell}
  \, \bar{q} \gamma_\mu \mathrm{P}_\mathrm{L}  q^\prime + \mathrm{h.c.}\right) \,,  \label{effective_Lagrangian_all}
\eer
where effective couplings to different quark fields $q$ are related as
\ber
c^b_\mathrm{R} \left(\mu \right)   &=&c^s_\mathrm{R} \left(\mu \right) = c^d_\mathrm{R} \left(\mu \right),\qquad c^c_\mathrm{R} \left(\mu \right) = c^u_\mathrm{R} \left(\mu \right), \\
c^s_\mathrm{L} \left(\mu \right)   &=& c^d_\mathrm{L} \left(\mu \right), \qquad \qquad \qquad c^c_\mathrm{L} \left(\mu \right) = c^u_\mathrm{L} \left(\mu \right),\\
  3 c_\mathrm{L}^u + 2 c_\mathrm{L}^{\nu_\mu e}
  &=& \sqrt{2}  \mathrm{G}_{u}, \qquad \quad ~~ - 3 c_\mathrm{L}^d + c_\mathrm{L}^{\nu_\mu e} = 2\sqrt{2}  \mathrm{G}_d,\\
  c_\mathrm{L}^{u } - c^u_\mathrm{R}
  &=& \sqrt{2} \mathrm{\tilde{G}}_u , \qquad \qquad \quad ~~~ c_\mathrm{L}^{d} - c^d_\mathrm{R} 
  = -\sqrt{2} \mathrm{\tilde{G}}_d \,,
\eer
with scale-independent generalizations of the Fermi constant determined up to order $\mathrm{O}(\alpha)$:
\ber
\mathrm{G}_u &=& 1.14570(23) \times 10^{-5}~\mathrm{GeV}^{-2}, \qquad
\mathrm{G}_d = 1.18211(21) \times 10^{-5}~\mathrm{GeV}^{-2}, \\
\mathrm{\tilde{G}}_u &=& 1.16841(20) \times 10^{-5}~\mathrm{GeV}^{-2}, \qquad
\mathrm{\tilde{G}}_d = 1.18154(21) \times 10^{-5}~\mathrm{GeV}^{-2}.
\eer

In Table~\ref{results_couplings_running}, we present the results for all effective couplings in the quark Lagrangian (\ref{effective_Lagrangian_all}) at $\mu = 2~\mathrm{GeV}$. The uncertainty of neutral current couplings comes mainly from the variation of the electroweak matching scale. The error of the charged current coupling $c^{q q'}$ is due to an unaccounted anomalous dimension of order $\alpha \alpha_s^2$. $V_{q q'}$ denotes a CKM matrix element.
\begin{table*}[h]
\centering
\caption{Effective couplings in $n_f = 4$ Fermi theory of neutrino-quark interaction at $\mu = 2~\mathrm{GeV}$.}
\label{results_couplings_running}
\begin{minipage}{\linewidth}  
\footnotesize
\centering
\begin{tabular}{|l|c|c|c|c|c|c|c|c|c|}
\hline          
& $ c^{u}_\mathrm{L}$ & $ c^{u}_\mathrm{R}$ & $ c^{d}_\mathrm{L}$ & $ c^{d}_\mathrm{R}$  & $c^{q q'} / V_{q q'} \left( M_Z\right)$ \\
\hline
$ \mu = 2~\mathrm{GeV} $ & $1.14065(10)$ &$-0.51173(37)$ &$-1.41478(10)$ &$0.25617(19)$ & $3.32685(8)$ \\
\hline
\end{tabular}
\end{minipage}
\end{table*}

Note that a scheme parameter $a$ enters the expression for $c^{q q'}$ coming from the one-loop matching condition on the effective field theory side as well as from the two-loop anomalous dimension. Performing Naive Dimensional Regularization (NDR), the relevant tensor product is expressed through the dimension of space-time $\mathrm{d}$ as~\cite{Buras:1989xd,Dugan:1990df,Herrlich:1994kh}
\ber
 \gamma^\alpha \gamma^\beta \gamma^\mu \mathrm{P}_\mathrm{L} \otimes \gamma_\mu \gamma_\beta \gamma_\alpha \mathrm{P}_\mathrm{L} = 4 \left( 1 + a \left( 4-\mathrm{d}\right) \right) \gamma^\mu \mathrm{P}_\mathrm{L} \otimes \gamma_\mu \mathrm{P}_\mathrm{L} + \mathrm{E} \left( a \right). \label{evanescent}
 \eer
We choose $a = -1$ so that the evanescent operator $\mathrm{E}$ projects to zero on the basis 
\ber
{\gamma^\mu\otimes \gamma_\mu,~\gamma^\mu \otimes \gamma_\mu \gamma_5,~\gamma_5 \gamma^\mu \otimes \gamma_\mu,~\gamma_5 \gamma^\mu \otimes \gamma_\mu \gamma_5 }.
\eer

Neutrino-electron scattering at energies of modern accelerator experiments and below is described by the theory with electron and neutrinos only. This work provides effective couplings in the interaction Lagrangian at a sub-percent level and presents absolute total cross section and energy spectra in neutrino-electron scattering quantifying errors for the first time. Hadronic contributions to the charge-isospin vector-vector correlation function provide the main source of uncertainty and require further investigations. Our cross section results can be useful to constrain the neutrino flux in modern and future neutrino experiments. For application to the broader program of neutrino-nucleon and neutrino-nucleus interactions, the neutrino-quark scattering Lagrangian with corresponding couplings is also presented at $\mu = 2~\mathrm{GeV}$ scale.

\section*{Acknowledgements}
I thank Richard J. Hill for reading this manuscript, useful discussions, 
and continuous support during this work. 
Work supported by the U.S. Department of Energy, Office of
Science, Office of High Energy Physics, under Award Number
DE-SC0019095. Fermilab is operated by Fermi Research Alliance, LLC
under Contract No. DE-AC02-07CH11359 with the United States Department
of Energy. The work of O. Tomalak was supported in part by the
Visiting Scholars Award Program of the Universities Research
Association. FeynCalc~\cite{Mertig:1990an,Shtabovenko:2016sxi},
LoopTools~\cite{Hahn:1998yk}, JaxoDraw~\cite{Binosi:2003yf},
Mathematica~\cite{Mathematica} and DataGraph were extremely useful in
this work.

\end{document}